\documentclass[11pt,twoside]{article}

\usepackage{asp2014}

\aspSuppressVolSlug
\resetcounters

\begin{document}

\title{Architecture of processing and analysis system for big astronomical data}

\author{Ivan~Kolosov$^1$, Sergey~Gerasimov$^1$, and Alexander~Meshcheryakov$^{2,3}$
\affil{$^1$Faculty of Computational Mathematics and Cybernetics of Lomonosov Moscow State University, Moscow, Russia; \email{zackwag32@gmail.com}}
\affil{$^2$Space Research Institute of Russian Academy of Sciences, Profsoyuznaya 84/32, 117997 Moscow, Russia; \email{mesch@iki.rssi.ru}}
\affil{$^3$Kazan Federal University, Kremlevskaya str.18, 420008 Kazan, Russia;}}

\paperauthor{Ivan~Kolosov}{zackwag32@gmail.com}{}{Lomonosov Moscow State University}{Faculty of Computational Mathematics and Cybernetics}{Moscow}{Moscow}{119991}{Russia}
\paperauthor{Sergey~Gerasimov}{sergun@gmail.com}{}{Lomonosov Moscow State University}{Faculty of Computational Mathematics and Cybernetics}{Moscow}{Moscow}{119991}{Russia}
\paperauthor{Alexander~Meshcheryakov}{mesch@iki.rssi.ru}{}{Russian Academy of Sciences}{Space Research Institute}{Moscow}{Moscow}{117997}{Russia}

\begin{abstract}
This work explores the use of big data technologies deployed in the cloud for processing of astronomical data. We have applied Hadoop and Spark to the task of co-adding astronomical images. We compared the overhead and execution time of these frameworks. We conclude that performance of both frameworks is generally on par. The Spark API is more flexible, which allows one to easily construct astronomical data processing pipelines.

\end{abstract}

\section{Introduction}
In the Big Data Era (see \citet{Zhang2015}) the progress in modern astronomy relies on the processing vast amounts of raw astronomical images produced by sky surveys. The ability of researchers to implement their own processing pipelines enables new scientific discoveries (see e.g. \citet{Lang2014,Koposov2015,Chilingarian2015}). But batch processing of large amounts of data (like \citet{Lang2014}) is a challenging task for a single astronomer or even a group of researchers due to lack of publicly available tools that are capable of such processing, lack of ready processing solutions that can span multiple processing steps and lack of control over computing infrastructure.

We appeal to Big Data technologies deployed in the cloud as a possible solution of these problems. Big Data technologies such as Apache Hadoop, Apache Spark are designed to process large amounts of data and scale with the number of computing nodes. Thus, they can be of interest to researchers that aim to solve their big data problems in astronomy.

The procedure of image co-addition takes a central place in astronomical data reduction pipeline, since this processing step substantially improves a quality of imaging data available for sky objects and drastically reduces the data volume. The use of Hadoop for image coaddition was explored in \citet{Wiley2011}. In this paper, we evaluate the use of successor of Hadoop Mapreduce, the Apache Spark cluster-computing framework, for the same task.

\section{Apache Hadoop}

Hadoop is a distributed computing framework optimized for fault-tolerant distributed processing of large amounts of data on clusters of computing nodes. The programming model of Hadoop exposes two operations called Map and Reduce that operate on pairs of keys and values. Map maps input key-value pairs to zero or more key-value pairs. Reduce reduces a set of key-value pairs that correspond to a single key into a key-value pair. Before doing Reduce, the framework performs a grouping operation called shuffle.

Hadoop uses a resource management system called YARN (Yet Another Resource Negotiator) that gives containers to applications. Each container has memory, CPU cores and other resources associated with it. Hadoop requests a container for each map or reduce task. The amount of memory and CPU cores given to mapper and reducer containers can be adjusted by the user.

\section{Apache Spark}

Spark is a data processing framework that is capable of caching data in memory. While Hadoop achieves fault-tolerant processing by storing intermediate output to disk, Spark uses its concept of Resilient Distributed Datasets (RDD). An RDD can be created from a data source or by transforming other RDDs. The user defines RDD transformations using Spark API, and Spark builds a computation graph and schedules its execution on the cluster. If a node of the graph fails, it is recomputed. These properties let Spark be faster for iterative algorithms as it avoids using the disk.

In Spark's computation model, a predefined number of executors is launched, each in its own container. An executor is then given tasks. This reduces overhead associated with launching tasks, but prohibits using different configurations for different kinds of tasks.

\section{Experiments on distributed image coaddition}

Our approach to distributed image coaddition involves tiling the region of interest with rectangular tiles. Tiles have an area of overlap needed to correctly process border objects. We project every input image onto every tile it belongs to, then group projected images by tile and coadd each group. This approach is easily programmed in both Hadoop and Spark. For our experiments (see below) we implemented the coaddition algorithm with Hadoop (in Java) and Spark (in Python). SWarp \cite{Bertin2002} was used to perform operations of background subtraction, reprojection and coaddition on astronomical images.

We used a popular set of SDSS Stripe 82 imaging data (r-band) for our experiments. 300 \(0.5 \times 0.5^{\circ}\) tiles were defined.  All tiles together made up a strip \(30^{\circ}\) wide by \(\alpha\) and \(2.5^{\circ}\) wide by \( \delta \) with its center in \( \alpha, \delta = 51.5912, 0.01311 \). Each tile had \(0.1^{\circ}\) wide overlap with neighbor tiles.

Microsoft Azure HDinsight cluster of 12 D12 type worker nodes was used for experiments. Each node had 28 GB RAM, 4 CPU cores and 200 GB local SSD\@. For Spark, we had to make sure that each executor had enough RAM for every computation stage. We ended up using 60 executors with 4736 MiB RAM each. For Hadoop jobs, we gave the same amount of memory to reducer tasks and 2048 MiB to Mapper tasks.

The SDSS data was stored in Microsoft Azure Blob Storage. We aggregated input images into bigger files, one file for each combination of SDSS run and camcol. This was done to make file lookup faster and enable using bigger input splits. The total volume of all data was \textasciitilde868 GB\@. We have used 128 MiB input splits.

For experiments with 868 GB input, we had to split each job in half and join the results of the two jobs. This was done because the size of temporary files exceeded the volume of local storage. All the metrics given for 868 GB jobs are made up by summing metrics for the three jobs that this job was split into.

\articlefigure{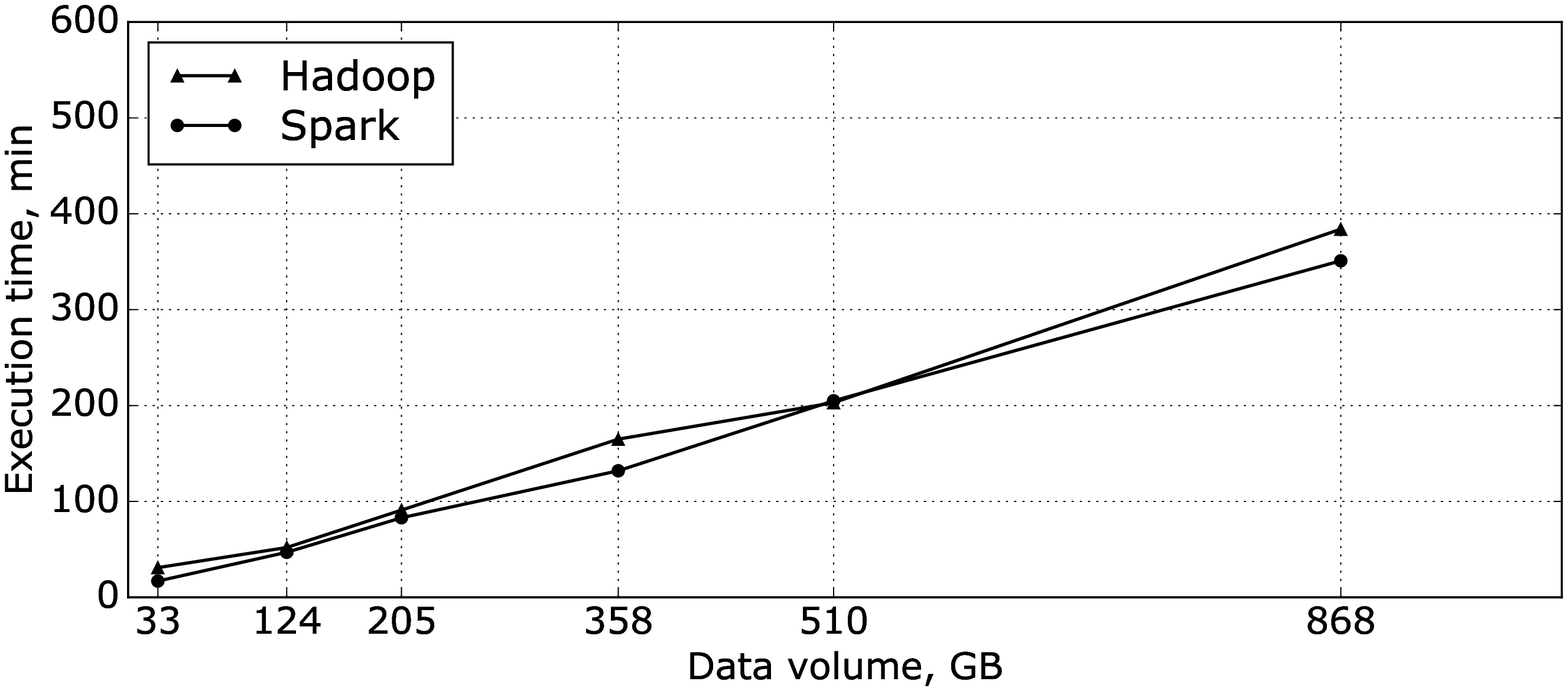}{fig-execution-time}{Execution time of Hadoop vs Spark}

To measure the overhead of both frameworks, we used event logs kept by the frameworks to compute execution time across all tasks, then used application logs to calculate user code execution time. The estimated framework overhead is equal to time across all tasks minus user code time divided by time across all tasks. We give a plot of these estimates on the left part of figure~\ref{fig-overhead-and-cpu}.

\articlefiguretwo{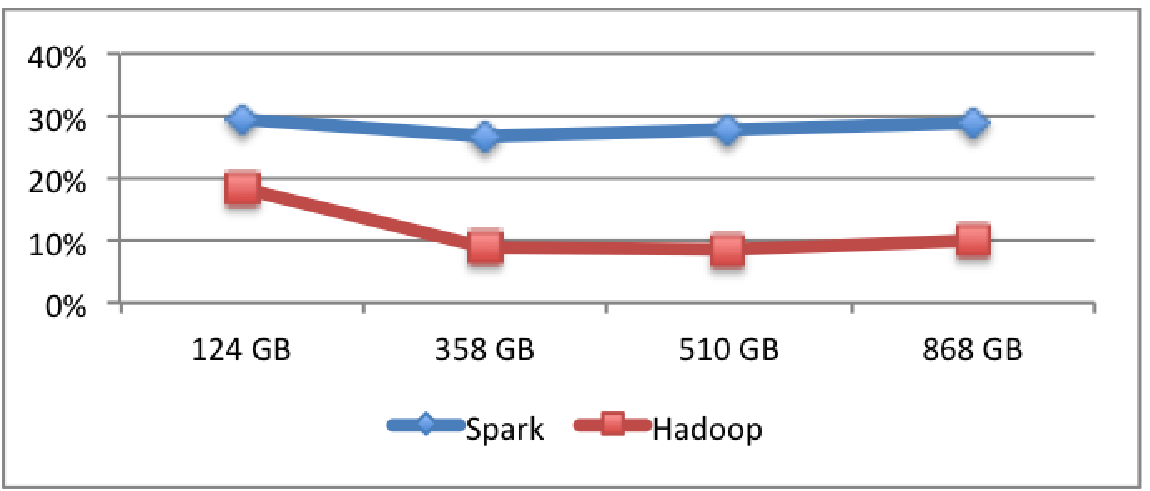}{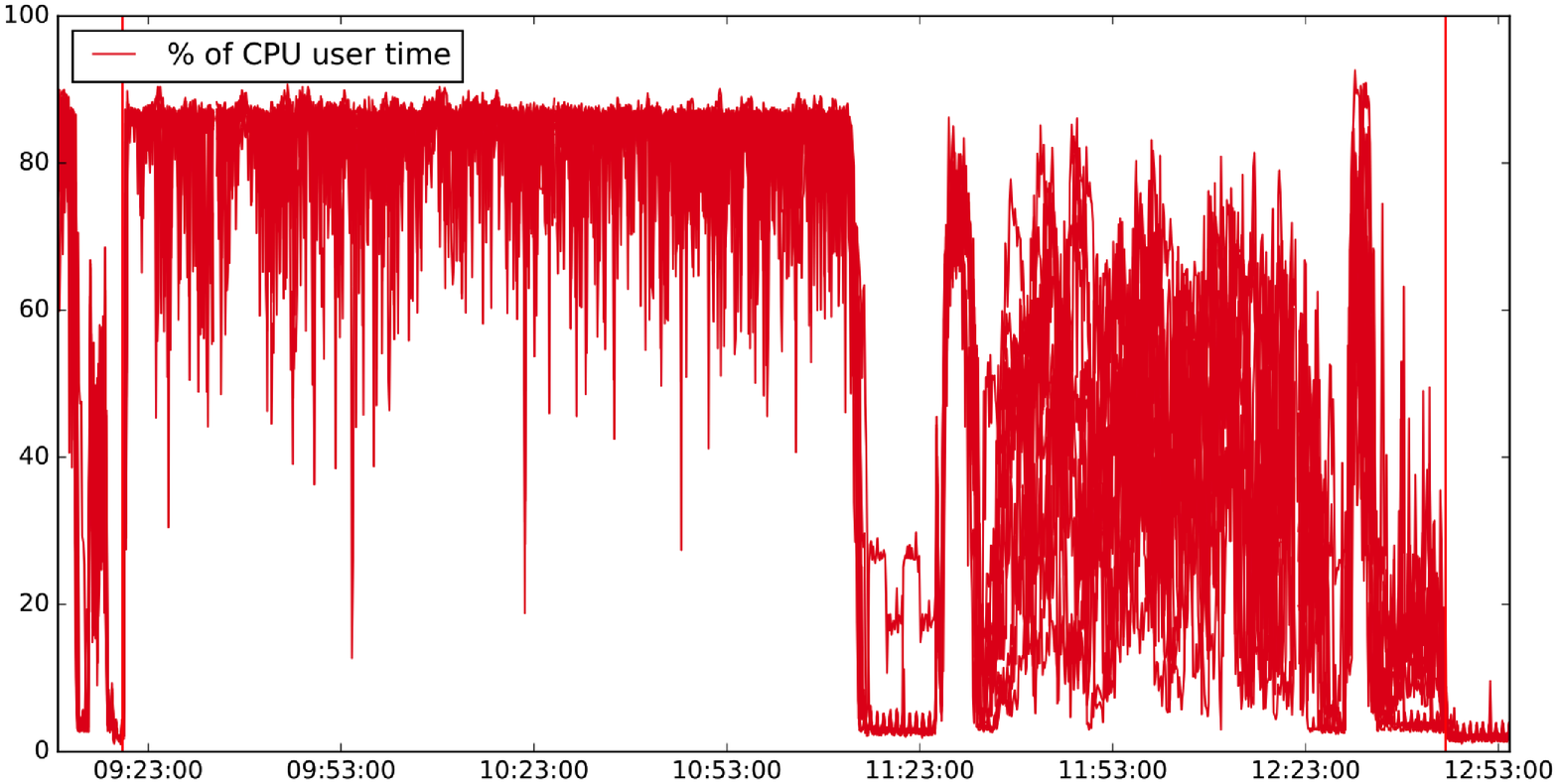}{fig-overhead-and-cpu}{\emph{Left}: Aggregate time spent executing framework code relative to overall aggregate time. \emph{Right}: CPU user time on each node, 510 GB on Spark}

We also measured the portion of CPU time spent executing userland code as a measure of how well cluster resources were utilized. Spark results are shown on figure~\ref{fig-overhead-and-cpu} on the right. Each line shows CPU time on one of the 12 nodes. Vertical lines show time boundaries of the job.

\section{Conclusion}

We co-added SDSS Stripe 82 images with Hadoop and Spark and compared the resulting performances of both frameworks. We conclude that performance of both frameworks is generally on par. Spark has more flexible API, which allows one to easily combine multiple reduction steps in the astronomical data processing pipeline. Hadoop, on the other hand, allows for more fine-grained configuration, e.g.\ it lets the user specify different configurations for various execution stages.

\acknowledgements This work is supported by Russian Foundation for Basic Research grants 14-22-03111-ofi-m and 14-22-03111-ofi-m. We also thank Microsoft Azure for Research for providing us with computing resources. AM acknowledges hospitality of the Kazan Federal University (KFU) and support by the Russian Government Program of Competitive Growth of KFU.

\bibliography{P1-36}  

\end{document}